# A general method for obtaining degenerate solutions to the Dirac and Weyl equations and a discussion on the experimental detection of degenerate states

Georgios N. Tsigaridas[1,*], Aristides I. Kechriniotis[2], Christos A. Tsonos[2] and Konstantinos K. Delibasis[3]

[1]Department of Physics, School of Applied Mathematical and Physical Sciences, National Technical University of Athens, GR-15780 Zografou Athens, Greece

[2]Department of Physics, University of Thessaly, GR-35100 Lamia, Greece

[3]Department of Computer Science and Biomedical Informatics, University of Thessaly, GR-35131 Lamia, Greece

[*]Corresponding Author. E-mail: gtsig@mail.ntua.gr

**Abstract**

In this work we describe a general method for obtaining degenerate solutions to the Dirac equation, corresponding to an infinite number of electromagnetic 4-potentials and fields, which are explicitly calculated. In more detail, using four arbitrary real functions, one can automatically construct a spinor which is solution to the Dirac equation for an infinite number of electromagnetic 4-potentials, defined by those functions. An interesting characteristic of these solutions is that, in the case of Dirac particles with non-zero mass, the degenerate spinors should be localized, both in space and time. Our method is also extended to the cases of massless Dirac and Weyl particles, where the localization of the spinors is no longer required. Finally, we propose two experimental methods for detecting the presence of degenerate states.

1. **Introduction**

We consider the Dirac equation in the form

$$i\gamma^\mu \partial_\mu \Psi + a_\mu \gamma^\mu \Psi - m\Psi = 0 \qquad (1)$$

where $\gamma^\mu$, $\mu = 0,1,2,3$, are the standard Dirac matrices, $m$ is the mass of the particle and $a_\mu = qU_\mu$ where $q$ is the electric charge of the particle and $U_\mu$ is the electromagnetic 4-potential. It should also be noted that Eq. (1) is written in natural units, where the speed of light in vacuum $c$ and the reduced Planck constant $\hbar$ are both set equal to one.

In a recent article [1] we have shown that all solutions to the Dirac equation satisfying the conditions $\Psi^\dagger \gamma \Psi = 0$ and $\Psi^T \gamma^2 \Psi \neq 0$, where $\gamma = \gamma^0 + i\gamma^1\gamma^2\gamma^3$, are degenerate, corresponding to an infinite number of electromagnetic 4-potentials which are explicitly calculated through Theorem 5.4. We have also shown that all solutions to the Weyl equations are degenerate. In this case, the corresponding electromagnetic 4-potentials are calculated through Theorem 3.1. In [2-5] we have extended these results providing several classes of degenerate solutions to the Dirac and Weyl equations for massive [2, 5] and massless [3, 4] particles, and describing their physical properties and potential applications. Furthermore, in [4] we discuss some very interesting properties of Weyl particles, mainly regarding their localization.

In this work, we provide a general method for obtaining degenerate solutions to the Dirac equation for real 4-potentials, which are explicitly calculated. The method is described in detail in section 2, and in section 3 is extended to massless Dirac and Weyl particles. In section 4 we discuss two experimental techniques for detecting the presence of degenerate states and the transition between these states and the non-degenerate ones. Our conclusions are presented in section 5. We have also added two appendices for providing the necessary mathematical background.

## 2. Formulation of the method and description of the degenerate spinors and the corresponding electromagnetic 4-potentials in the case of massive Dirac particles

It is easy to verify that any spinor of the form

$$\Psi = T \begin{pmatrix} \cos\varphi \\ 1-\sin\varphi \\ \cos\varphi \\ 1-\sin\varphi \end{pmatrix} + R \begin{pmatrix} -\cos\varphi \\ 1+\sin\varphi \\ \cos\varphi \\ -1-\sin\varphi \end{pmatrix} \quad (2)$$

where $T, R$ are arbitrary complex functions of the spatial coordinates and time and $\varphi \neq n\pi + \pi/2$, $n \in \mathbb{Z}$ is an arbitrary real constant is degenerate.

Substituting the spinor given by Eq. (2) into the Dirac equation, we obtain the following system of equations

$$(\cos\varphi \partial_1 + \sin\varphi \partial_3 + \partial_0) R = i(a_1 \cos\varphi + a_3 \sin\varphi + a_0) R \quad (3)$$

$$(\cos\varphi \partial_1 + \sin\varphi \partial_3 + \partial_0) T = i(a_1 \cos\varphi + a_3 \sin\varphi + a_0) T \quad (4)$$

$$(-i\cos\varphi \partial_2 - \partial_3 - \sin\varphi \partial_0) R = (a_2 \cos\varphi - ia_3 - ia_0 \sin\varphi) R + im(1-\sin\varphi) T \quad (5)$$

$$(-i\cos\varphi \partial_2 + \partial_3 + \sin\varphi \partial_0) T = (a_2 \cos\varphi + ia_3 + ia_0 \sin\varphi) T + im(1+\sin\varphi) R \quad (6)$$

Defining the matrix

$$\Pi = \begin{pmatrix} \cos\varphi & 0 & 0 & 0 \\ 0 & -i\cos\varphi & -i\cos\varphi & 0 \\ \sin\varphi & -1 & 1 & 0 \\ 1 & -\sin\varphi & \sin\varphi & 1 \end{pmatrix} \qquad (7)$$

and setting

$$\begin{pmatrix} D_1 \\ D_2 \\ D_3 \\ D_0 \end{pmatrix} = \Pi^T \begin{pmatrix} \partial_1 \\ \partial_2 \\ \partial_3 \\ \partial_0 \end{pmatrix} \qquad (8)$$

where $\Pi^T$ is the transpose of $\Pi$, the system of equations (3)-(6) can be written as

$$D_1 R = A_1 R \qquad (9)$$

$$D_1 T = A_1 T \qquad (10)$$

$$D_2 R = A_2 R + im(1-\sin\varphi)T \qquad (11)$$

$$D_3 T = A_3 T + im(1+\sin\varphi)R \qquad (12)$$

where

$$A_1 = i(a_1 \cos\varphi + a_3 \sin\varphi + a_0) \qquad (13)$$

$$A_2 = a_2 \cos\varphi - ia_3 - ia_0 \sin\varphi \qquad (14)$$

$$A_3 = a_2 \cos\varphi + ia_3 + ia_0 \sin\varphi \qquad (15)$$

As shown in Appendix A, using the following transformation of the coordinates $x_0, x_1, x_2, x_3$

$$\begin{pmatrix} x_1 \\ x_2 \\ x_3 \\ x_0 \end{pmatrix} = \Pi \begin{pmatrix} s_1 \\ s_2 \\ s_3 \\ s_0 \end{pmatrix} \qquad (16)$$

the linear differential operators $D_i$, $i=1,2,3$ can be written as $\partial/\partial s_i = \tilde{\partial}_i$. Consequently, the system of equations (9)-(12) takes the form

$$\tilde{\partial}_1 \tilde{R} = \tilde{A}_1 \tilde{R} \qquad (17)$$

$$\tilde{\partial}_1\tilde{T} = \tilde{A}_1\tilde{T} \tag{18}$$

$$\tilde{\partial}_2\tilde{R} = \tilde{A}_2\tilde{R} + im(1-\sin\varphi)\tilde{T} \tag{19}$$

$$\tilde{\partial}_3\tilde{T} = \tilde{A}_3\tilde{T} + im(1+\sin\varphi)\tilde{R} \tag{20}$$

where $\tilde{A}_1, \tilde{A}_2, \tilde{A}_3, \tilde{R}, \tilde{T}$ are the functions $A_1, A_2, A_3, R, T$ expressed in the coordinates $s_0, s_1, s_2, s_3$.

Multiplying equations (17), (18) with $\exp\left(-\int \tilde{A}_1 ds_1\right)$ we obtain that

$$\tilde{\partial}_1\left(\tilde{R}\exp\left(-\int \tilde{A}_1 ds_1\right)\right) = 0 \tag{21}$$

$$\tilde{\partial}_1\left(\tilde{T}\exp\left(-\int \tilde{A}_1 ds_1\right)\right) = 0 \tag{22}$$

Consequently, the functions $\tilde{R}, \tilde{T}$ can be written as

$$\tilde{R} = \exp\left(\int \tilde{A}_1 ds_1\right)\tilde{g}_R \tag{23}$$

$$\tilde{T} = \exp\left(\int \tilde{A}_1 ds_1\right)\tilde{g}_T \tag{24}$$

where $\tilde{g}_R, \tilde{g}_T$ are arbitrary complex functions of the coordinates $s_0, s_2, s_3$. Substituting equations (23), (24) into (19), (20) and supposing that $\tilde{A}_1$ depends only on $s_0, s_1$ we obtain the following system of equations for the functions $\tilde{g}_R, \tilde{g}_T$:

$$\left(\tilde{\partial}_2 - \tilde{A}_2\right)\tilde{g}_R = im(1-\sin\varphi)\tilde{g}_T \tag{25}$$

$$\left(\tilde{\partial}_3 - \tilde{A}_3\right)\tilde{g}_T = im(1+\sin\varphi)\tilde{g}_R \tag{26}$$

Multiplying Eq. (25) by $im(1+\sin\varphi)$ and Eq. (26) by $im(1-\sin\varphi)$, yields that

$$\left(\tilde{\partial}_2 - \tilde{A}_2\right)\left(im(1+\sin\varphi)\tilde{g}_R\right) = -m^2\cos^2\varphi\,\tilde{g}_T \tag{27}$$

$$\left(\tilde{\partial}_3 - \tilde{A}_3\right)\left(im(1-\sin\varphi)\tilde{g}_T\right) = -m^2\cos^2\varphi\,\tilde{g}_R \tag{28}$$

which, according to equations (25), (26), can be written as

$$\left(\tilde{\partial}_2 - \tilde{A}_2\right)\left(\tilde{\partial}_3 - \tilde{A}_3\right)\tilde{g}_T = -m^2\cos^2\varphi\,\tilde{g}_T \tag{29}$$

$$\left(\tilde{\partial}_3 - \tilde{A}_3\right)\left(\tilde{\partial}_2 - \tilde{A}_2\right)\tilde{g}_R = -m^2 \cos^2\varphi\, \tilde{g}_R \qquad (30)$$

Multiplying equations (29), (30) by $\exp\left(-\int \tilde{A}_2 ds_2 - \int \tilde{A}_3 ds_3\right)$ and assuming that $\tilde{\partial}_3 \tilde{A}_2 = 0$ and $\tilde{\partial}_2 \tilde{A}_3 = 0$, the above system of equations takes the following form:

$$\tilde{\partial}_2 \tilde{\partial}_3 \left(\exp\left(-\int \tilde{A}_2 ds_2 - \int \tilde{A}_3 ds_3\right)\tilde{g}_T\right) = -m^2 \cos^2\varphi\, \exp\left(-\int \tilde{A}_2 ds_2 - \int \tilde{A}_3 ds_3\right)\tilde{g}_T \qquad (31)$$

$$\tilde{\partial}_2 \tilde{\partial}_3 \left(\exp\left(-\int \tilde{A}_2 ds_2 - \int \tilde{A}_3 ds_3\right)\tilde{g}_R\right) = -m^2 \cos^2\varphi\, \exp\left(-\int \tilde{A}_2 ds_2 - \int \tilde{A}_3 ds_3\right)\tilde{g}_R \qquad (32)$$

Consequently, the functions $\tilde{g}_R$, $\tilde{g}_T$ can be written as

$$\tilde{g}_R = \exp\left(\int \tilde{A}_2 ds_2 + \int \tilde{A}_3 ds_3\right)\tilde{W}_R \qquad (33)$$

$$\tilde{g}_T = \exp\left(\int \tilde{A}_2 ds_2 + \int \tilde{A}_3 ds_3\right)\tilde{W}_T \qquad (34)$$

where $\tilde{W}_R(s_0, s_2, s_3)$, $\tilde{W}_T(s_0, s_2, s_3)$ are solutions to the differential equation

$$\tilde{\partial}_2 \tilde{\partial}_3 \tilde{W} = -m^2 \cos^2\varphi\, \tilde{W} \qquad (35)$$

Here, we have also assumed that $\tilde{\partial}_1 \tilde{A}_2 = 0$ and $\tilde{\partial}_1 \tilde{A}_3 = 0$, because the functions $\tilde{g}_R, \tilde{g}_T$ depend only on $s_0, s_2, s_3$.

Thus, assuming that

$$\tilde{\partial}_2 \tilde{A}_1 = 0,\ \tilde{\partial}_3 \tilde{A}_1 = 0,\ \tilde{\partial}_1 \tilde{A}_2 = 0,\ \tilde{\partial}_3 \tilde{A}_2 = 0,\ \tilde{\partial}_1 \tilde{A}_3 = 0,\ \tilde{\partial}_2 \tilde{A}_3 = 0 \qquad (36)$$

the functions $\tilde{R}, \tilde{T}$ can be written as

$$\tilde{R} = \exp\left(\int \tilde{A}_1 ds_1 + \int \tilde{A}_2 ds_2 + \int \tilde{A}_3 ds_3\right)\tilde{W}_R \qquad (37)$$

$$\tilde{T} = \exp\left(\int \tilde{A}_1 ds_1 + \int \tilde{A}_2 ds_2 + \int \tilde{A}_3 ds_3\right)\tilde{W}_T \qquad (38)$$

Finally, substituting the above expressions into Eq. (20), yields that the functions $\tilde{W}_T, \tilde{W}_R$ should be related through the following formula:

$$\tilde{\partial}_3 \tilde{W}_T = im(1 + \sin\varphi)\tilde{W}_R \qquad (39)$$

Thus, any spinor of the form

$$\tilde{\Psi} = \exp\left(\int \tilde{A}_1 ds_1 + \int \tilde{A}_2 ds_2 + \int \tilde{A}_3 ds_3\right)$$
$$\times \left( im(1+\sin\varphi)\int \tilde{W} ds_3 \begin{pmatrix} \cos\varphi \\ 1-\sin\varphi \\ \cos\varphi \\ 1-\sin\varphi \end{pmatrix} + \tilde{W} \begin{pmatrix} -\cos\varphi \\ 1+\sin\varphi \\ \cos\varphi \\ -1-\sin\varphi \end{pmatrix} \right) \quad (40)$$

where $\tilde{A}_1, \tilde{A}_2, \tilde{A}_3$ satisfy the conditions given by Eq. (36), and $\tilde{W}(s_0, s_2, s_3)$ is an arbitrary solution to the differential equation (35), is degenerate solution to the Dirac equation.

An interesting remark is that, according to equations (13)-(15), assuming that the 4-potentials $(a_0, a_1, a_2, a_3)$ are real, the function $A_1$ becomes imaginary and the function $A_3$ becomes the complex conjugate of $A_2$. Thus, the 4-potentials $(a_0, a_1, a_2, a_3)$ are given by the formulae

$$a_0 = h \quad (41)$$

$$a_1 = -h\cos\varphi + \text{Im}(A_1)\sec\varphi + \text{Im}(A_2)\tan\varphi \quad (42)$$

$$a_2 = \text{Re}(A_2)\sec\varphi \quad (43)$$

$$a_3 = -h\sin\varphi - \text{Im}(A_2) \quad (44)$$

where $h$ is an arbitrary real function of the spatial coordinates and time, $\text{Im}(A_1)$, $\text{Im}(A_2)$ are the imaginary parts of $A_1$, $A_2$, respectively, and $\text{Re}(A_2)$ is the real part of $A_2$. Here, it should be mentioned that the functions $A_1, A_2, A_3$, and consequently the 4-potentials $(a_0, a_1, a_2, a_3)$ could also depend on the mass of the particles.

Another interesting remark is that the coordinates $s_0, s_1$ are real functions of the coordinates $x_1, x_2, x_3, x_0$, as it can be easily verified from the matrix $\Pi$. Consequently, assuming that $\tilde{A}_2$, $\tilde{A}_3$ depend only on $s_0$ and defining the real functions $\tilde{f}_{1I}(s_0, s_1)$, $\tilde{f}_{2R}(s_0)$, $\tilde{f}_{2I}(s_0)$ through the formulae

$$\tilde{f}_{1I}(s_0, s_1) = -i\tilde{A}_1 \tag{45}$$

$$\tilde{f}_{2R}(s_0) = (\tilde{A}_2 + \tilde{A}_3)/2 \tag{46}$$

$$\tilde{f}_{2I}(s_0) = -i(\tilde{A}_2 - \tilde{A}_3)/2 \tag{47}$$

it is easy to verify that the spinor

$$\tilde{\Psi} = \exp\left(i\int \tilde{f}_{1I}(s_0, s_1)ds_1 + \tilde{f}_{2R}(s_0)(s_2 + s_3) + i\tilde{f}_{2I}(s_0)(s_2 - s_3)\right)$$

$$\times \left( im(1+\sin\varphi)\int \tilde{W}ds_3 \begin{pmatrix} \cos\varphi \\ 1-\sin\varphi \\ \cos\varphi \\ 1-\sin\varphi \end{pmatrix} + \tilde{W}\begin{pmatrix} -\cos\varphi \\ 1+\sin\varphi \\ \cos\varphi \\ -1-\sin\varphi \end{pmatrix} \right) \tag{48}$$

is a degenerate solution to the Dirac equation, for the real 4-potentials given by the following expressions:

$$a_0 = h \tag{49}$$

$$a_1 = -h\cos\varphi + f_{1I}\sec\varphi + f_{2I}\tan\varphi \tag{50}$$

$$a_2 = f_{2R}\sec\varphi \tag{51}$$

$$a_3 = -h\sin\varphi - f_{2I} \tag{52}$$

Here, $f_{1I}, f_{2R}, f_{2I}$ are real functions of the spatial coordinates and time, connected to the functions $\tilde{f}_{1I}(s_0, s_1)$, $\tilde{f}_{2R}(s_0)$, $\tilde{f}_{2I}(s_0)$ through the following transformation of the coordinates:

$$\begin{pmatrix} s_1 \\ s_2 \\ s_3 \\ s_0 \end{pmatrix} = \Pi^{-1} \begin{pmatrix} x_1 \\ x_2 \\ x_3 \\ x_0 \end{pmatrix} \tag{53}$$

where $\Pi^{-1}$ is the inverse matrix of $\Pi$, given by the formula

$$\Pi^{-1} = \begin{pmatrix} \sec\varphi & 0 & 0 & 0 \\ \frac{1}{2}\tan\varphi & \frac{1}{2}\sec\varphi & -\frac{1}{2} & 0 \\ -\frac{1}{2}\tan\varphi & \frac{i}{2}\sec\varphi & \frac{1}{2} & 0 \\ -\cos\varphi & 0 & -\sin\varphi & 1 \end{pmatrix} \tag{54}$$

Consequently, for any combination of the arbitrary real functions $\tilde{f}_{1I}(s_0,s_1)$, $\tilde{f}_{2R}(s_0)$, $\tilde{f}_{2I}(s_0)$, one can automatically construct the spinor given by Eq. (48), which is degenerate solution to the Dirac equation for the infinite number of real 4-potentials given by the formulae (49)-(52). Finally, using the coordinate transformation described by Eq. (16), the degenerate spinor (48) can be expressed in terms of the spatial and temporal coordinates $x_0, x_1, x_2, x_3$.

Additionally, according to Theorem 5.4 in [1], the spinor given by Eq. (48) will also be solution to the Dirac equation for the 4-potentials

$$b_\mu = a_\mu + s\kappa_\mu, \quad \mu = 0,1,2,3 \tag{55}$$

where

$$(\kappa_0, \kappa_1, \kappa_2, \kappa_3) = \left(1, -\frac{\Psi^T \gamma^0 \gamma^1 \gamma^2 \Psi}{\Psi^T \gamma^2 \Psi}, -\frac{\Psi^T \gamma^0 \Psi}{\Psi^T \gamma^2 \Psi}, \frac{\Psi^T \gamma^0 \gamma^2 \gamma^3 \Psi}{\Psi^T \gamma^2 \Psi}\right) \tag{56}$$
$$= (1, -\cos\varphi, 0, -\sin\varphi)$$

and $s$ is an arbitrary real function of the spatial coordinates and time. It is evident that the 4-potentials $b_\mu$ coincide with the 4-potentials $a_\mu$, given by equations (49)-(52).

The electromagnetic fields (in Gaussian units) corresponding to the 4-potentials $a_\mu$, or $b_\mu$, can be easily calculated through the formulae [6, 7]

$$\mathbf{E} = -\nabla U - \frac{\partial \mathbf{A}}{\partial t}, \qquad \mathbf{B} = \nabla \times \mathbf{A} \tag{57}$$

where $U = a_0/q$ is the electric potential and $\mathbf{A} = -(1/q)(a_1\mathbf{i} + a_2\mathbf{j} + a_3\mathbf{k})$ is the magnetic vector potential. It should also be noted that in the above formulae the speed of light has been set equal to one, since we are working in the natural system of units, where $\hbar = c = 1$.

Furthermore, as shown in Appendix B, the differential equation (35) has solutions of the form

$$\tilde{W}(s_0, s_2, s_3) = g(s_0)\exp\left(-\frac{m^2 \cos^2 \varphi}{k(s_0)}s_2 + k(s_0)s_3\right) \tag{58}$$

where $g(s_0), k(s_0) \neq 0$ are arbitrary complex functions of $s_0$. In the following, for simplicity, we assume that $k(s_0)$ is constant. Using the above expression for $\tilde{W}(s_0, s_2, s_3)$, the spinor given by Eq. (48) becomes

$$\tilde{\Psi} = \exp\left(i\int \tilde{f}_{1I}(s_0, s_1)ds_1 + \tilde{f}_{2R}(s_0)(s_2 + s_3) + i\tilde{f}_{2I}(s_0)(s_2 - s_3)\right)$$
$$\times g(s_0)\exp\left(-\frac{m^2 \cos^2 \varphi}{k}s_2\right)\exp(ks_3) \tag{59}$$
$$\times \left(i\frac{m(1+\sin\varphi)}{k}\begin{pmatrix}\cos\varphi \\ 1-\sin\varphi \\ \cos\varphi \\ 1-\sin\varphi\end{pmatrix} + \begin{pmatrix}-\cos\varphi \\ 1+\sin\varphi \\ \cos\varphi \\ -1-\sin\varphi\end{pmatrix}\right)$$

Another interesting remark is that, according to the transformation given by Eq. (53), the coordinates $s_1, s_2, s_3, s_0$ can be written as

$$s_1 = x\sec\varphi \tag{60}$$

$$s_2 = \frac{1}{2}x\tan\varphi + \frac{i}{2}y\sec\varphi - \frac{1}{2}z \tag{61}$$

$$s_3 = -\frac{1}{2}x\tan\varphi + \frac{i}{2}y\sec\varphi + \frac{1}{2}z \tag{62}$$

$$s_0 = t - x\cos\varphi - z\sin\varphi \tag{63}$$

where we have also made the substitution $x_1 \to x$, $x_2 \to y$, $x_3 \to z$, $x_0 \to t$. Consequently,

$$s_2 + s_3 = iy\sec\varphi \tag{64}$$

and

$$s_2 - s_3 = x\tan\varphi + z \tag{65}$$

Additionally, it is evident that the coordinates $s_1, s_0$ are real function of $x, y, z, t$. Therefore, the factor

$$\exp\left(i\int \tilde{f}_{1I}(s_0, s_1)ds_1 + \tilde{f}_{2R}(s_0)(s_2 + s_3) + i\tilde{f}_{2I}(s_0)(s_2 - s_3)\right) \tag{66}$$

in the spinors given by equations (48), (59) is a real function of the coordinates and the 4-potentials. Thus, all the information regarding the 4-potentials is incorporated into the phase of the spinor.

Furthermore, the factor

$$g(s_0)\exp\left(-\frac{m^2\cos^2\varphi}{k}s_2\right)\exp(ks_3) \tag{67}$$

in Eq. (59), in terms of the coordinates $x, y, z, t$, takes the following form:

$$g(t-x\cos\varphi-z\sin\varphi)\exp\left(-\frac{m^2\cos^2\varphi+k^2}{2k}(x\tan\varphi+z)\right)$$
$$\times\exp\left(-i\frac{m^2\cos^2\varphi-k^2}{2k}y\sec\varphi\right) \tag{68}$$

Consequently, it the spinor given by Eq. (59) tends to infinity, as $x, z$ tend to infinity. To overcome this problem, we must use an appropriate form for the arbitrary function $g(t-x\cos\varphi-z\sin\varphi)$, e.g.

$$g(t-x\cos\varphi-z\sin\varphi) = c_1\exp\left(-k_R(t-x\cos\varphi-z\sin\varphi)^2\right)$$
$$\times\exp\left(-ik_I(t-x\cos\varphi-z\sin\varphi)\right) \tag{69}$$

where $c_1$ is an arbitrary complex constant, $k_R$ is an arbitrary real and positive constant and $k_I$ is an arbitrary real constant. In the above expression we have also introduced the factor $\exp(-ik_I(t-x\cos\varphi-z\sin\varphi))$ to ensure the wave-nature of the spinor.

As an example, we consider the special case that $\tilde{f}_{1I}(s_0,s_1) = k_1 s_0$, $\tilde{f}_{2R}(s_0) = k_2 s_0$, $\tilde{f}_{2I}(s_0) = k_3 s_0$, where $k_1, k_2, k_3$ are arbitrary real constants. Then, the 4-potentials given by equations (49)-(52) take the following form:

$$a_0 = h \tag{70}$$

$$a_1 = -h\cos\varphi - (k_1 + k_2\sin\varphi)(x - t\sec\varphi + z\tan\varphi) \tag{71}$$

$$a_2 = -k_3(x + z\tan\varphi - t\sec\varphi) \tag{72}$$

$$a_3 = -h\sin\varphi + k_2(x\cos\varphi + z\sin\varphi - t) \tag{73}$$

According to Eq. (57), the electromagnetic fields corresponding to these 4-potentials are given by the formulae

$$\mathbf{E} = \left(k_1 \sec\varphi + k_2 \tan\varphi - \cos\varphi \frac{\partial h_q}{\partial t} - \frac{\partial h_q}{\partial x}\right)\mathbf{i}$$
$$+ \left(k_3 \sec\varphi - \frac{\partial h_q}{\partial y}\right)\mathbf{j} - \left(k_2 + \sin\varphi \frac{\partial h_q}{\partial t} + \frac{\partial h_q}{\partial z}\right)\mathbf{k} \tag{74}$$

$$\mathbf{B} = \left(k_2 \sec\varphi + k_1 \tan\varphi + \cos\varphi \frac{\partial h_q}{\partial z} - \sin\varphi \frac{\partial h_q}{\partial x}\right)\mathbf{j}$$
$$+ \left(-k_3 \tan\varphi + \sin\varphi \frac{\partial h_q}{\partial y}\right)\mathbf{i} + \left(k_3 - \cos\varphi \frac{\partial h_q}{\partial y}\right)\mathbf{k} \tag{75}$$

where $h_q = h/q$.

Furthermore, according to Eq. (59), the spinor that is solution to the Dirac equation for the above 4-potentials, takes the following form:

$$\Psi = c_1 \exp\left(i\sec\varphi\left((k_1 + k_2 \sin\varphi)x + k_3 y - k_2 z \cos\varphi\right)(t - x\cos\varphi - z\sin\varphi)\right)$$
$$\times \exp\left(-k_R(t - x\cos\varphi - z\sin\varphi)^2\right)\exp\left(-ik_I(t - x\cos\varphi - z\sin\varphi)\right)$$
$$\times \exp\left(-\frac{m^2\cos^2\varphi + k^2}{2k}(x\tan\varphi + z)\right)\exp\left(-i\frac{m^2\cos^2\varphi - k^2}{2k}y\sec\varphi\right)$$
$$\times \left(i\frac{m(1+\sin\varphi)}{k}\begin{pmatrix}\cos\varphi\\1-\sin\varphi\\\cos\varphi\\1-\sin\varphi\end{pmatrix} + \begin{pmatrix}-\cos\varphi\\1+\sin\varphi\\\cos\varphi\\-1-\sin\varphi\end{pmatrix}\right) \tag{76}$$

where we have also considered the coordinate transformation given by Eq. (53).

Thus, particles described by the degenerate spinor given by Eq. (76) can exist in the same quantum state in the wide variety of electromagnetic fields described by equations (74), (75).

An important remark is that particles described by the above spinors are localized, both in space and time, since $\lim_{x\to\pm\infty}\Psi = \lim_{z\to\pm\infty}\Psi = \lim_{t\to\pm\infty}\Psi = 0$. This practically means that the family of degenerate solutions for massive Dirac particles presented in this article, describes particles in localized states. As it will be shown in the following section, this is no longer required in the cases of massless Dirac and Weyl particles.

Another characteristic of these solutions is that the expected values of the projections of the spin of the particles along the x, y, and z axes, as calculated by the following formulae [8, 9]:

$$S_x = \frac{i}{2}\Psi^\dagger \gamma^2 \gamma^3 \Psi = \frac{1}{2}\cos\varphi \left(\frac{m^2(1+\cos 2\varphi) - 2k^2}{m^2(1+\cos 2\varphi) + 2k^2}\right)|\Psi|^2 \tag{77}$$

$$S_y = \frac{i}{2}\Psi^\dagger \gamma^3 \gamma^1 \Psi = 0 \tag{78}$$

$$S_z = \frac{i}{2}\Psi^\dagger \gamma^1 \gamma^2 \Psi = \frac{1}{2}\sin\varphi \left(\frac{m^2(1+\cos 2\varphi) - 2k^2}{m^2(1+\cos 2\varphi) + 2k^2}\right)|\Psi|^2 \tag{79}$$

are functions of the mass of the particles and the modulus of the spinor, defined as $|\Psi|^2 = \Psi^\dagger \Psi$, which is also a function of the mass of the particles and the spatial and temporal coordinates. However, setting $k = m\cos\varphi$, the expected values of the projections of the spin of the particles along the x, y, and z axes become all equal to zero.

### 3. Extension of the method to massless Dirac and Weyl particles

In the special case that the mass of the particles becomes zero, it can be easily verified through equations (19), (20), (23), (24), (33) and (34) that a degenerate solution to the massless Dirac equation for the real 4-potentials given by equations (49)-(51) is the following:

$$\tilde{\Psi} = \exp\left(i\int \tilde{f}_{1I}(s_0, s_1)ds_1 + \tilde{f}_{2R}(s_0)(s_2 + s_3) + i\tilde{f}_{2I}(s_0)(s_2 - s_3)\right)$$

$$\times \left(\tilde{W}_T(s_0, s_2)\begin{pmatrix}\cos\varphi \\ 1-\sin\varphi \\ \cos\varphi \\ 1-\sin\varphi\end{pmatrix} + \tilde{W}_R(s_0, s_3)\begin{pmatrix}-\cos\varphi \\ 1+\sin\varphi \\ \cos\varphi \\ -1-\sin\varphi\end{pmatrix}\right) \tag{80}$$

where $\tilde{W}_T(s_0, s_2)$, $\tilde{W}_R(s_0, s_3)$ are arbitrary complex functions of the coordinates $s_0, s_2$ and $s_0, s_3$ respectively. Obviously, special care must be taken to ensure that the spinor given by Eq. (80) is bound for all values of the spatial and temporal coordinates. The simplest choice satisfying this condition is setting the functions $\tilde{W}_T(s_0, s_2)$, $\tilde{W}_R(s_0, s_3)$ as follows:

$$\tilde{W}_T(s_0, s_2) = c_T \exp(-ik_I(t - x\cos\varphi - z\sin\varphi)) \tag{81}$$

$$\tilde{W}_R(s_0, s_2) = c_R \exp(-ik_I(t - x\cos\varphi - z\sin\varphi)) \tag{82}$$

where we have used the term $\exp(-ik_I(t-x\cos\varphi-z\sin\varphi))$ to ensure the wave-nature of the spinor. Here, $c_T, c_R$ are arbitrary complex constants. As an example, we consider the following spinor

$$\Psi = \exp\left(i\sec\varphi\left((k_1+k_2\sin\varphi)x + k_3 y - k_2 z\cos\varphi\right)(t-x\cos\varphi-z\sin\varphi)\right)$$
$$\times \exp(-ik_I(t-x\cos\varphi-z\sin\varphi))\left[c_T\begin{pmatrix}\cos\varphi\\1-\sin\varphi\\\cos\varphi\\1-\sin\varphi\end{pmatrix} + c_R\begin{pmatrix}-\cos\varphi\\1+\sin\varphi\\\cos\varphi\\-1-\sin\varphi\end{pmatrix}\right] \quad (83)$$

which is degenerate solution to the massless Dirac equation for the real 4-potentials given by equations (70)-(73), corresponding to the electromagnetic fields given by equations (74), (75). Thus, a massless Dirac particle described by the above spinor will exist in the same quantum state in the wide variety of electromagnetic fields given by equations (74), (75).

An important remark is that, contrary to the case of massive particles, there is no need to impose spatial or temporal restrictions for massless particles, which are free to move in all space and time. Additionally, in the case of massless particles, the expected values of the projections of the spin along the x, y, and z axes become all constants, taking the following values:

$$S_x = \frac{i}{2}\Psi^\dagger\gamma^2\gamma^3\Psi = -2\cos\varphi\left(|c_R|^2 - |c_T|^2 + \left(|c_T|^2 + |c_R|^2\right)\sin\varphi\right) \quad (84)$$

$$S_y = \frac{i}{2}\Psi^\dagger\gamma^3\gamma^1\Psi = 0 \quad (85)$$

$$S_z = \frac{i}{2}\Psi^\dagger\gamma^1\gamma^2\Psi = -2\sin\varphi\left(|c_R|^2 - |c_T|^2 + \left(|c_T|^2 + |c_R|^2\right)\sin\varphi\right) \quad (86)$$

Furthermore, it is important to mention that, in the case of $\tilde{W}_R(s_0,s_3)=0$ or $\tilde{W}_T(s_0,s_2)=0$, the degenerate spinors given by Eq. (80) take the form $\tilde{\Psi} = (\tilde{\psi}_T, \tilde{\psi}_T)^T$ or $\tilde{\Psi} = (\tilde{\psi}_R, -\tilde{\psi}_R)^T$ respectively, where

$$\tilde{\psi}_T = \exp\left(i\int \tilde{f}_{1I}(s_0,s_1)ds_1 + \tilde{f}_{2R}(s_0)(s_2+s_3) + i\tilde{f}_{2I}(s_0)(s_2-s_3)\right)$$
$$\times \tilde{W}_T(s_0,s_2)\begin{pmatrix}\cos\varphi\\1-\sin\varphi\end{pmatrix} \quad (87)$$

and

$$\tilde{\psi}_R = \exp\left(i\int \tilde{f}_{1I}(s_0, s_1) ds_1 + \tilde{f}_{2R}(s_0)(s_2 + s_3) + i\tilde{f}_{2I}(s_0)(s_2 - s_3)\right)$$
$$\times \tilde{W}_R(s_0, s_3) \begin{pmatrix} -\cos\varphi \\ 1 + \sin\varphi \end{pmatrix} \tag{88}$$

According to Theorem 3.1 in [1], the spinors $\tilde{\psi}_T$ are solutions to the Weyl equation in the form

$$i\sigma^\mu \partial_\mu \Psi + a_\mu \sigma^\mu \Psi = 0 \tag{89}$$

corresponding to particles with positive helicity and the spinors $\tilde{\psi}_R$ are solutions to the Weyl equation in the form

$$i\sigma^\mu \partial_\mu \Psi - 2i\sigma^0 \partial_0 \Psi + a_\mu \sigma^\mu \Psi - 2a_0 \sigma^0 \Psi = 0 \tag{90}$$

corresponding to particles with negative helicity. Here, $\sigma^\mu$ are the standard Pauli matrices, $a_\mu = qA_\mu$, $q$ is the electric charge of the particles and $A_\mu$ is the electromagnetic 4-potential, as in the case of the Dirac equation. The Weyl equations are also expressed in natural units, where $\hbar = c = 1$.

As an example, we consider the spinors

$$\psi_T = \exp\left(i\sec\varphi\left((k_1 + k_2\sin\varphi)x + k_3 y - k_2 z\cos\varphi\right)(t - x\cos\varphi - z\sin\varphi)\right)$$
$$\times \exp(-ik_I(t - x\cos\varphi - z\sin\varphi)) \begin{pmatrix} \cos\varphi \\ 1 - \sin\varphi \end{pmatrix} \tag{91}$$

which are solutions to the Weyl equation for particles with positive helicity and the spinors

$$\psi_R = \exp\left(i\sec\varphi\left((k_1 + k_2\sin\varphi)x + k_3 y - k_2 z\cos\varphi\right)(t - x\cos\varphi - z\sin\varphi)\right)$$
$$\times \exp(-ik_I(t - x\cos\varphi - z\sin\varphi)) \begin{pmatrix} -\cos\varphi \\ 1 + \sin\varphi \end{pmatrix} \tag{92}$$

which are solutions to the Weyl equation for particles with negative helicity. In both cases, the 4-potentials corresponding to these solutions are given by equations (70) – (73).

An interesting remark is that the phase factor

$$\exp\left(i\int \tilde{f}_{1I}(s_0, s_1) ds_1 + \tilde{f}_{2R}(s_0)(s_2 + s_3) + i\tilde{f}_{2I}(s_0)(s_2 - s_3)\right) \tag{93}$$

containing the information regarding the electromagnetic 4-potentials is the same in all cases, namely for massive Dirac particles, massless Dirac particles and Weyl particles.

Furthermore, according to Theorem 3.1 in [1], all Weyl spinors are degenerate, corresponding to the 4-potentials

$$b_{\mu 1} = a_{\mu 1} + s\kappa_{\mu 1} \tag{94}$$

where

$$(\kappa_{01}, \kappa_{11}, \kappa_{21}, \kappa_{31}) = \left(1, -\frac{\psi_T^\dagger \sigma^1 \psi_T}{\psi_T^\dagger \psi_T}, -\frac{\psi_T^\dagger \sigma^2 \psi_T}{\psi_T^\dagger \psi_T}, -\frac{\psi_T^\dagger \sigma^3 \psi_T}{\psi_T^\dagger \psi_T}\right)$$
$$= (1, -\cos\varphi, 0, -\sin\varphi) \tag{95}$$

in the case of particles with positive helicity, and

$$b_{\mu 2} = a_{\mu 2} + s\kappa_{\mu 2} \tag{96}$$

where

$$(\kappa_{02}, \kappa_{12}, \kappa_{22}, \kappa_{32}) = \left(1, \frac{\psi_R^\dagger \sigma^1 \psi_R}{\psi_R^\dagger \psi_R}, \frac{\psi_R^\dagger \sigma^2 \psi_R}{\psi_R^\dagger \psi_R}, \frac{\psi_R^\dagger \sigma^3 \psi_R}{\psi_R^\dagger \psi_R}\right)$$
$$= (1, -\cos\varphi, 0, -\sin\varphi) \tag{97}$$

in the case of particles with negative helicity. It is evident that, in both cases, the 4-potentials, and consequently the electromagnetic fields, are the same to those corresponding to Dirac particles.

Thus, for any combination of the arbitrary real functions $\tilde{f}_{1I}(s_0, s_1)$, $\tilde{f}_{2R}(s_0)$, $\tilde{f}_{2I}(s_0)$ one can automatically construct spinors which are degenerate solutions to the massive Dirac, massless Dirac and Weyl equations, for the infinite number of 4-potentials given by equations (49)-(52).

### 4. On the experimental detection of degenerate states

In the special that the function $h$ depends only on time and $k_1 = k_2 = k_3 = 0$, the electromagnetic fields, given by equations (74) and (75), take the simple form

$$\mathbf{E} = -\frac{\partial h_q}{\partial t}(\cos\varphi \mathbf{i} + \sin\varphi \mathbf{k}), \quad \mathbf{B} = \mathbf{0}. \tag{98}$$

Additionally, in the above case, the spinors defined by equations (76), (83), and (91), (92) describe particles moving parallel to the vector $\cos\varphi \mathbf{i} + \sin\varphi \mathbf{k}$. Consequently, the state of the particles will not be affected by the presence of an electric field of arbitrary time dependence, applied along their direction of motion. This practically means that the electric current transferred by charged particles in degenerate states will not change if a voltage, of arbitrary magnitude and time dependence, is applied along the direction of motion of the particles. This behavior can be easily detected experimentally in materials supporting massless Dirac or Weyl particles, such as graphene sheets and Weyl semimetals [10-15].

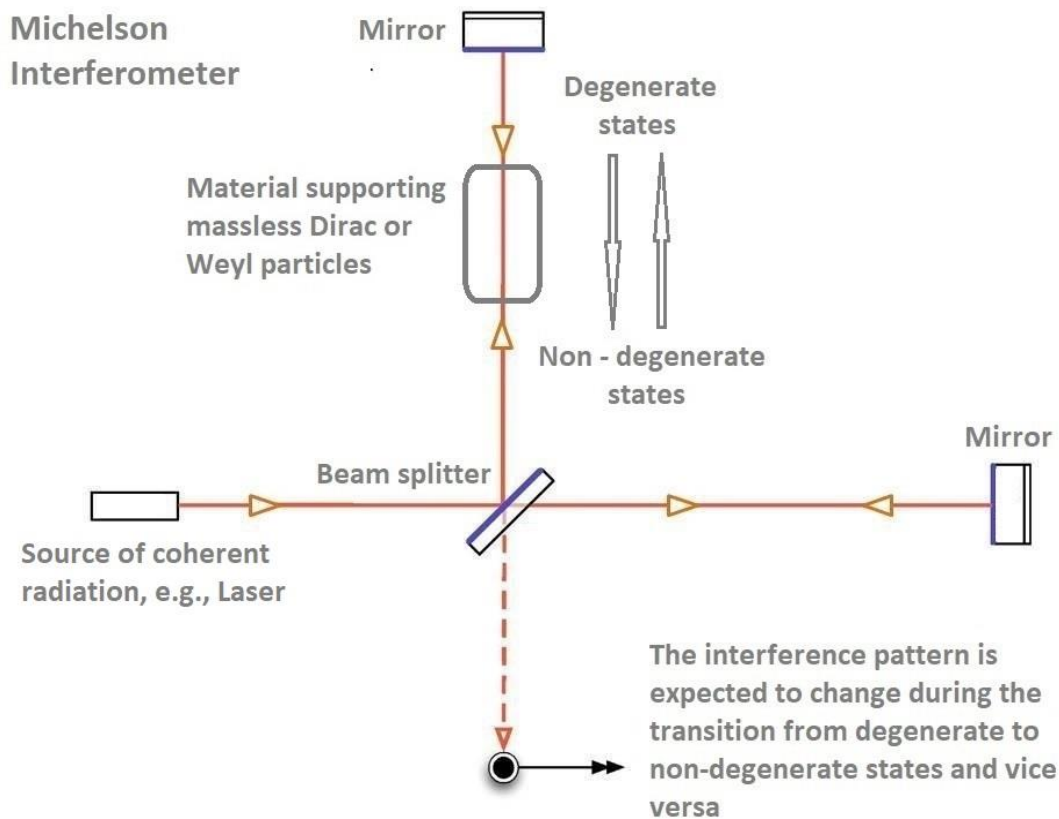

**Figure 1**: A proposed method for experimentally detecting the transition between degenerate and non-degenerate states using a Michelson interferometer.

Furthermore, as shown in [3], the state of Weyl and massless Dirac particles described by degenerate spinors, will not be affected by the presence of a plane electromagnetic wave, e.g., a laser beam of arbitrary polarization, propagating along the direction of motion of the particles. Thus, particles in degenerate states and electromagnetic waves can propagate along the same direction without interacting with each other, which obviously is not the case for charged particles in non-degenerate states.

This result can be used to detect the presence of degenerate states through an interferometric method. In more detail, if we place a material where charged particles are in degenerate states in one arm of a Michelson interferometer, the electromagnetic wave propagating through this arm will behave as if it was propagating in vacuum. On the other hand, if the particles are not in degenerate states, they will interact with the electromagnetic wave affecting its velocity and consequently its phase. Therefore, the transition from non-degenerate to degenerate states and vice-versa could be easily detected through the changes in the interference pattern produced by the interferometer, as shown in figure 1.

It should be noted that not all particles are expected to move parallel to the electromagnetic wave and consequently some of them will interact with the wave, even in the degenerate states. However, interferometric experiments are so sensitive, that even if a small fraction of the particles stops interacting with the electromagnetic wave, it should be sufficient to induce a measurable change in the diffraction pattern.

Thus, the techniques described above can be used to detect the presence of degenerate states, as well as the transition between degenerate and non-degenerate states. In addition, these methods can be used for determining if quantum states corresponding to massless Dirac or Weyl particles are excited in a material. More details on these techniques, as well as the potential applications of our theory in various fields of physics involving the interaction of charged particles with electromagnetic fields, will be provided in future works.

## 5. Conclusions

In conclusion, we have provided a general method for obtaining degenerate solutions to the Dirac equation, corresponding to an infinite number of electromagnetic 4-potentials and fields, which are explicitly calculated. The electromagnetic 4-potenials are constructed using four arbitrary real functions, three of which appear in the phase of the degenerate spinors, corresponding to these 4-potentials. Furthermore, the method is extended to the cases of massless Dirac and Weyl particles, where the information regarding the 4-potentials is also encoded in the phase of the spinors. Additionally, an important remark is that, in the case of massive Dirac particles, the degenerate spinors describe localized states, which is no longer required for massless Dirac and Weyl particles. Finally, we describe two experimental methods for detecting the presence of degenerate states, as well as the transition between degenerate and non-degenerate states.

**Appendix A**

We suppose that $A = (a_{ij}) \in Mat_n(C)$ is an invertible $n \times n$ matrix. We also define the linear differential operators $D_i, i = 1, ..., n$ through the formula

$$\begin{pmatrix} D_1 \\ . \\ . \\ . \\ D_n \end{pmatrix} = A^T \begin{pmatrix} \partial_1 \\ . \\ . \\ . \\ \partial_n \end{pmatrix} \qquad (A1)$$

where $\partial_i = \dfrac{\partial}{\partial x_i}$. Then, using the following linear transformation of the coordinates

$$\begin{pmatrix} x_1 \\ . \\ . \\ . \\ x_n \end{pmatrix} = A \begin{pmatrix} s_1 \\ . \\ . \\ . \\ s_n \end{pmatrix} \qquad (A2)$$

we can verify that

$$D_i = \tilde{\partial}_i, i = 1,...,n \qquad (A3)$$

where $\tilde{\partial}_i = \dfrac{\partial}{\partial s_i}$. Indeed, defining the matrix

$$B = (b_{ij}) = A^{-1} \qquad (A4)$$

and using Eq. (A2) we obtain that

$$s_i = \sum_{k=1}^{n} b_{ik} x_k , \; i = 1,...,n \qquad (A5)$$

Then, using equations (A1), (A4), (A5), it is easy to verify that, for any function $U(x_1,...x_n) = \tilde{U}(s_1,...,s_n)$, the following is true:

$$\begin{pmatrix} D_1 \\ . \\ . \\ . \\ D_n \end{pmatrix} U = A^T \begin{pmatrix} \partial_1 U \\ . \\ . \\ . \\ \partial_n U \end{pmatrix} = A^T \begin{pmatrix} \sum_{i=1}^{n} \tilde{\partial}_i \tilde{U} \partial_1 s_i \\ . \\ . \\ . \\ \sum_{i=1}^{n} \tilde{\partial}_i \tilde{U} \partial_n s_i \end{pmatrix}$$

$$= A^T \begin{pmatrix} \sum_{i=1}^{n} b_{i1} \tilde{\partial}_i \tilde{U} \\ . \\ . \\ . \\ \sum_{i=1}^{n} b_{in} \tilde{\partial}_i \tilde{U} \end{pmatrix} = A^T B^T \begin{pmatrix} \tilde{\partial}_1 \tilde{U} \\ . \\ . \\ . \\ \tilde{\partial}_n \tilde{U} \end{pmatrix} = \begin{pmatrix} \tilde{\partial}_1 \\ . \\ . \\ . \\ \tilde{\partial}_n \end{pmatrix} \tilde{U}$$

(A6)

## Appendix B

We consider the differential equation

$$\tilde{\partial}_2 \tilde{\partial}_3 \tilde{W} = -m^2 \cos^2 \varphi \ \tilde{W} \tag{B1}$$

where $\tilde{W}$ is an arbitrary complex function of $s_0, s_2, s_3$. Assuming that $\tilde{W}$ can be written in the form

$$\tilde{W}(s_0, s_2, s_3) = \tilde{W}_2(s_0, s_2) \tilde{W}_3(s_0, s_3) \tag{B2}$$

the differential equation (B1) takes the form

$$\tilde{\partial}_2 \tilde{W}_2 \tilde{\partial}_3 \tilde{W}_3 = -m^2 \cos^2 \varphi \ \tilde{W}_2 \tilde{W}_3 \tag{B3}$$

Under the condition that

$$\frac{\tilde{\partial}_3 \tilde{W}_3}{\tilde{W}_3} = k(s_0) \tag{B4}$$

the solution of Eq. (B3) for $\tilde{W}_2$ is

$$\tilde{W}_2(s_0, s_2) = g_2(s_0) \exp\left(-\frac{m^2 \cos^2 \varphi}{k(s_0)} s_2\right) \tag{B5}$$

where $g_2(s_0), k(s_0) \neq 0$ are arbitrary complex functions of $s_0$. Additionally, the solution of Eq. (B4) for $\tilde{W}_3$ is

$$\tilde{W}_3(s_0, s_3) = g_3(s_0) \exp(k(s_0) s_3) \tag{B6}$$

where $g_3(s_0)$ is an arbitrary complex function of $s_0$.

Thus, the solution of Eq. (B1) for $\tilde{W}$ can be expressed as

$$\tilde{W}(s_0, s_2, s_3) = g(s_0) \exp\left(-\frac{m^2 \cos^2 \varphi}{k(s_0)} s_2 + k(s_0) s_3\right) \tag{B7}$$

where $g(s_0) = g_2(s_0) g_3(s_0)$ is an arbitrary complex function of $s_0$.